\documentclass[a4paper]{jpconf}
\usepackage{graphicx}
\begin{document}
\title{Comparative study of discrete Boltzmann model and Navier-Stokes}

\author{Yudong Zhang$^{1,2}$, Aiguo Xu$^{1,3,*}$ and Guangcai Zhang$^{1}$}

\address{1 Laboratory of Computational Physics, Institute of Applied Physics and Computational Mathematics, Beijing, 100088, China

2 Key Laboratory of Transient Physics, Nanjing University of Science and Technology, Nanjing 210094, China

3 Center for Applied Physics and Technology, MOE Key Center for High Energy Density Physics Simulations, College of Engineering, Peking University, Beijing 100871, China
}

\ead{Xu\_Aiguo@iapcm.ac.cn}

\begin{abstract}
Discrete Boltzmann model (DBM) is a type of coarse-grained mesoscale kinetic model derived from the Boltzmann equation. Physically, it is roughly equivalent to a hydrodynamic model supplemented by a coarse-grained model for the relevant thermodynamic non-equilibrium (TNE) behaviours. The Navier-Stokes (NS) model is a traditional macroscopic hydrodynamic model based on continuity hypothesis and conservation laws. In this study, the two models are compared from two aspects, physical capability and computational cost, by simulating two kinds of flow problems including the thermal Couette flow and a Mach 3 step problem. In the cases where the TNE effects are weak, both the two models give accurate results for the hydrodynamic behaviour. Besides, DBM can provide more detailed non-equilibrium information, while the NS is more efficient if concern only the density, momentum, energy and their derived quantities. It is concluded that, if the TNE effects are strong or are to be investigated, the NS is insufficient while DBM is a good choice. While in the cases where the TNE effects are weak and only the macro flow fields are to be studied, the NS is more preferable.
\end{abstract}

\section{Introduction}
In recent years, discrete Boltzmann model (DBM) has been well developed to investigate various of complex flow situations where the thermodynamic non-equilibrium effect (TNE) is  significant~\cite{Xu2015Progess,Xu2016Progess,Xu2018Book}. It provides a new point of view to study the non-equilibrium characteristics of flow system. Up to now, DBM has been successfully used in non-equilibrium phase transition and multiphase flow~\cite{Gan2015Discrete}, flow instability~\cite{Lai2016Nonequilibrium,Feng2016Viscosity,Lin2017PRE}, reactive flow~\cite{Lin2017SR,Lin2018CF}, shock and  detonation~\cite{Xu2015PRE,Lin2016Double,Zhang2016Kinetic}, etc.

Compared with traditional hydrodynamic model, the Navier-Stokes (NS) equation~\cite{Book-Chen2002,Book-Fu2006}, DBM has more kinetic information. Generally, there are at least three different aspects between the two models. Firstly, the evolution equations of DBM are a set of the discrete particle velocity distribution, while the evolution equations of NS are those for the conservation laws of mass, momentum, and energy of control volume. Secondly, apart from the discretization of time and space, the discretization of particle velocity space is necessary for DBM. As a result, the actual number of evolution equations of DBM is determined by the discrete velocity model and usually is more than that of the NS. Thirdly, the numerical stability of DBM needs also consider the discrete velocities besides the general Courant$-$Friedrichs$-$Lewy (CFL) condition~\cite{Book-Fu2006}. Nevertheless, DBM is necessary in the following two cases: (i) the TNE is so strong that the NS does not work well any more, or (ii) not only the hydrodynamic behavior but also the relevant TNE effect are to be investigated.

Since each model has its own advantages and limitations. In this work, a comparative study of DBM and NS is made. Two kinds of problems, including thermal Couette flow and Mach 3 step problem, are simulated. The two models are both used under the same conditions. The physical capabilities and computation times are compared.

\section{Model and equations}
\subsection{Discrete Boltzmann model}
The evolution equation of DBM read
\begin{equation}\label{DBM1}
\frac{{\partial {f_i}}}{{\partial t}} + {v_{i\alpha }} \cdot \frac{{\partial {f_i}}}{{\partial {r_\alpha }}} =  - \frac{1}{\tau }({f_i} - {f_i}^ + ) \mathrm{,}
\end{equation}
where $f^+_i$ can be chosen as Maxwell distribution, Ellipsoidal Statistical (ES) distribution, and Shakhov distribution which correspond to BGK, ES-BGK, and S-model, respectively. According to the modeling of DBM, the discrete distribution function can be solved from the kinetic moments~\cite{Gan2013Lattice},
\begin{equation}\label{DBM2}
f_i^ +  = C_{ij}^{ - 1}{\hat f_j}   \mathrm{,}
\end{equation}
where $C_{ij}^{ - 1}$ is the coefficient matrix of discrete velocity and ${\hat f_j}$ is kinetic moment calculated from integral of the original distribution function.

The NTE quantities are defined as~\cite{Xu2015Progess}
\begin{equation}\label{NTE1}
\Delta _m^* = \sum\limits_i {({f_i} - f_i^{eq})\underbrace {({v_{i{\alpha _1}}} - {u_{i{\alpha _1}}}) \cdots ({v_{i{\alpha _m}}} - {u_{i{\alpha _m}}})}_m}   \mathrm{,}
\end{equation}
where $f_i^{eq}$ is discrete Maxwell distribution.
\subsection{Navier-Stokes equations}
Navier-Stokes equations include continuity equation, momentum conservation equation, and energy conservation which read~\cite{Book-Chen2002}
\begin{equation}\label{NS1}
\frac{{\partial \rho }}{{\partial t}} + \frac{{\partial (\rho {u_\alpha })}}{{\partial {r_\alpha }}} = 0  \mathrm{,}
\end{equation}
\begin{equation}\label{NS2}
\frac{{\partial \left( {\rho {u_\alpha }} \right)}}{{\partial t}} + \frac{\partial }{{\partial {r_\beta }}}\left[ {\rho (T{\delta _{\alpha \beta }} + {u_\alpha }{u_\beta }) - 2\mu \frac{{\partial {u_{ < \alpha }}}}{{\partial {r_{\beta  > }}}}} \right] = 0    \mathrm{,}
\end{equation}
\begin{equation}\label{NS3}
\frac{{\partial \left[ {\rho (e + \frac{{{u^2}}}{2})} \right]}}{{\partial t}} + \frac{\partial }{{\partial {r_\alpha }}}\left[ {\rho {u_\alpha }\left( {e + T + \frac{{{u^2}}}{2}} \right) - \left( {\kappa '\frac{{\partial T}}{{\partial {r_\alpha }}} + 2{u_\alpha }\mu \frac{{\partial {u_{ < \alpha }}}}{{\partial {r_{\beta  > }}}}} \right)} \right] = 0    \mathrm{,}
\end{equation}
where $\mu  = \tau \rho T$ and $\kappa ' = {c_p}\tau \rho T$ are coefficients of viscosity and heat flux, respectively. In fact, Navier-Stokes equation can also be derived from Eq.(1) by Champan Enskog expansion~\cite{Book-Chapman1970} under the first order approximation. The viscosity and heat flux are correspond to non-organizational momentum flux(NOMF) and non-organizational energy flux(NOEF)~\cite{Zhang2016Kinetic}, respectively.

\section{Results and Analysis}
\subsection{Thermal Couette flow}
Firstly, the thermal Couette flow is simulated by using DBM and numerical solving NS equations. The same time and space steps are adopted for two models. The space derivative is solved by second order upwind scheme in DBM and second center scheme in NS. The results are shown in Fig.1. Figure 1(a) shows the distribution of velocity at various time. Figure 1(b) gives the distribution of temperature when the flow reaches to its steady state. The results of two models are almost identical and both fit well with analysis solution.
\begin{figure}
  \centering
  \includegraphics[width=0.7\textwidth]{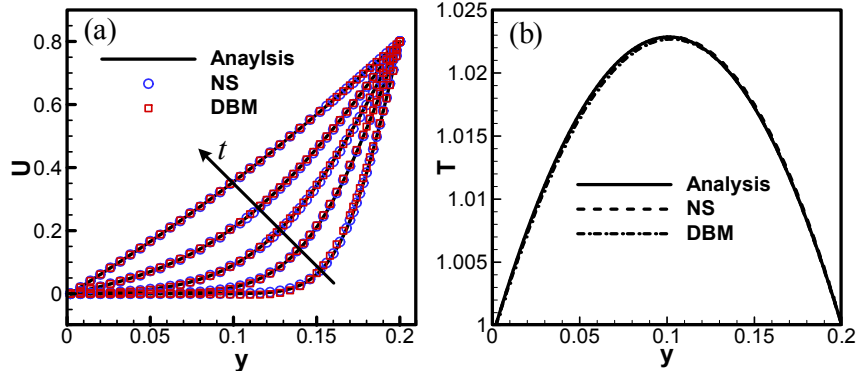}\\
  \caption{(a) The distribution of velocity over time; (b) The distribution of temperature (steady-state).}\label{fig1}
\end{figure}
\subsection{Mach 3 step problem}
As a classical two dimensional compressible flow problem, the Mach 3 step problem is widely used to verify various new models~\cite{Woodward1984The}. Here, we use DBM and NS simulate this problem under same condition. The NND scheme~\cite{Zhang1988NON} is adopted to solve space derivative in both model. Figure 2 shows the contour maps of density and pressure by two models. The results are well consistent with each other.
\begin{figure}
  \centering
  \includegraphics[width=0.8\textwidth]{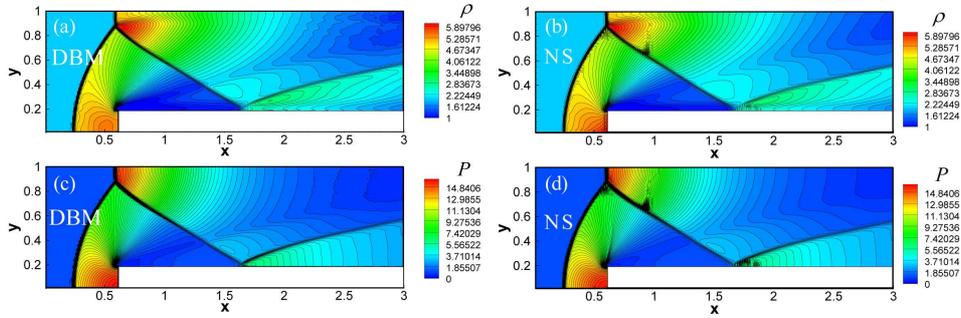}\\
  \caption{Contour maps of density and pressure: (a) density by DBM, (b) density by NS, (c) pressure by DBM, (d) pressure by NS.}\label{fig2}
\end{figure}
In addition, the NTE quantities provided by DBM are also shown in Fig. 3, which shows the non-equilibrium characteristics of the whole flow field. Four kinds NTE are displayed and each describes the non-equilibrium strength from different perspective. Those NTE quantities cannot obtained by NS.
\begin{figure}
  \centering
  \includegraphics[width=0.8\textwidth]{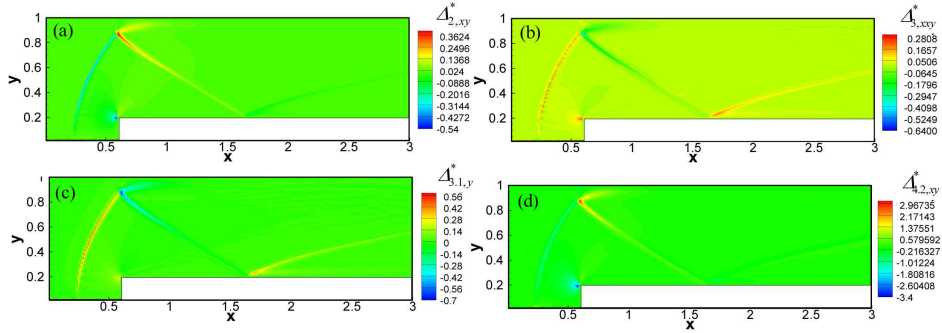}\\
  \caption{Contour maps of TNE quantities provided by DBM. (a)$\Delta _{2,xy}^*$, (b)$\Delta _{3,xxy}^*$, (c)$\Delta _{3.1,y}^*$, (d)$\Delta _{4.2,xy}^*$.}\label{fig3}
\end{figure}
\subsection{Computational efficiency comparison}
From the previous two example, we have learned that both DBM and NS can well describe thermal flows and high speed compressible flows. Now, the computational efficiencies of the two models are compared. The computational mesh for thermal Couette flow and Mach 3 step problem are $4\times 100$ and $300\times 100$ , respectively. The computation times are shown in Table \ref{Tab1}.The calculation is performed on a PC with Inter(R) Core(TM) i7-4790 CPU @3.60GHz

Obviously, NS has a higher efficiency than DBM. So, if the NS can meet the requirements, it is better to choose NS. However, it should be note that DBM is easier to program than NS.
\begin{table}[h]
\centering
\caption{\label{tabone}Computational time comparison between DBM and NS.}\label{Tab1}
\begin{center}
\lineup
\begin{tabular}{*{3}{ccc}}
\br
$\0\0$&Thermal Couette flow&Mach 3 step\cr
\mr
DBM (s/1000 steps)&17.08  &506.1 \cr
NS (s/1000 steps) &3.14    &227.8 \cr
\br
\end{tabular}
\end{center}
\end{table}

\section{Conclusion}
The comparative study of DBM and NS is made by two kinds of flow problems. In the cases where both the two models are suitable, the simulation results of the two models are well inconsistent with each other. But when it comes to computational efficiency, NS has an advantage over DBM if only the evolutions of generally hydrodynamic quantities (density, flow velocity, temperature, pressure and their derived variables) are to be investigated. Physically, the DBM can provide more detailed non-equilibrium information than the NS. In the cases where the TNE effects are strong or the TNE effects are to be investigated, the NS fails to work and the DBM is physically necessary. In the cases where the TNE effects are weak and one needs only the macro flow fields, the NS is more preferred for saving computational time, especially when simulating high dimensional problems.

\ack{Acknowledgement}
The work is supported by National Natural Science Foundation of China [under grant nos. 11475028 and 11772064] and Science Challenge Project (under Grant No. JCKY2016212A501 and TZ2016002).

\section*{References}
\bibliography{iopart-num}

\end{document}